\documentclass[runningheads]{llncs}
\usepackage[T1]{fontenc}
\usepackage{graphicx}
\usepackage{booktabs}
\usepackage{array}
\usepackage{caption}
\usepackage{xcolor}
\usepackage{hyperref}
\usepackage{amsmath}
\usepackage{amssymb}
\usepackage{tikz}
\usepackage{pgfplots}
\pgfplotsset{compat=1.18}
\usepackage{colortbl}
\usepackage{tcolorbox}
\usepackage{makecell}
\usepackage{multirow}

\definecolor{coordhier}{RGB}{180,80,120}
\definecolor{coorddebate}{RGB}{80,120,200}
\definecolor{coordconf}{RGB}{60,160,100}
\definecolor{coordcomp}{RGB}{200,130,50}
\definecolor{coordseq}{RGB}{120,120,120}
\definecolor{coordnone}{RGB}{160,80,80}
\definecolor{finmembg}{RGB}{255,235,235}
\definecolor{warnbg}{RGB}{255,243,220}
\definecolor{warnborder}{RGB}{200,150,50}
\definecolor{scorered}{RGB}{180,40,40}
\definecolor{scoreamber}{RGB}{180,120,0}
\definecolor{mutedgray}{RGB}{130,130,130}

\newcommand{\yes}{\cellcolor{green!15}\textcolor{green!50!black}{\textbf{\checkmark}}}
\newcommand{\no}{\cellcolor{red!10}\textcolor{red!70!black}{\textbf{$\times$}}}
\newcommand{\scoremid}[1]{\textcolor{scoreamber}{\textbf{#1}}}
\newcommand{\scorelow}[1]{\textcolor{scorered}{\textbf{#1}}}

\newif\ifcomments
\commentstrue
\ifcomments\newcommand{\comments}[1]{#1}\else\newcommand{\comments}[1]{}\fi

\begin{document}

\title{Toward Reliable Evaluation of LLM-Based Financial Multi-Agent Systems: Taxonomy, Coordination Primacy, and Cost Awareness}
\titlerunning{Toward Reliable Evaluation of LLM-Based Financial Multi-Agent Systems}

\author{Phat Nguyen\inst{1}\orcidID{0009-0004-0671-2487} \and
Thang Pham \inst{2}\orcidID{0000-0003-3984-641X}}
\authorrunning{P. Nguyen and T. Pham}
%
\institute{Georgia Institute of Technology, Atlanta GA 30332, USA \\
\email{cherry.07.skr@gmail.com} \and
Adobe Inc., San Jose CA 95110, USA \\
\email{thanpham@adobe.com}
}

\maketitle

\begin{abstract}
Multi-agent systems based on large language models (LLMs) for financial trading have grown rapidly since 2023, yet the field lacks a shared framework for understanding what drives performance or for evaluating claims credibly. This survey makes three contributions. First, we introduce a four-dimensional taxonomy, covering architecture pattern, coordination mechanism, memory architecture, and tool integration; applied to 12 multi-agent systems and two single-agent baselines. Second, we formulate the \textit{Coordination Primacy Hypothesis} (CPH): inter-agent coordination protocol design is a primary driver of trading decision quality, often exerting greater influence than model scaling. CPH is presented as a falsifiable research hypothesis supported by tiered structural evidence rather than as an empirically validated conclusion; its definitive validation requires evaluation infrastructure that does not yet exist in the field. Third, we document five pervasive evaluation failures (look-ahead bias, survivorship bias, backtesting overfitting, transaction cost neglect, and regime-shift blindness) and show that these can reverse the sign of reported returns. Building on the CPH and the evaluation critique, we introduce the \textit{Coordination Breakeven Spread} (CBS), a metric for determining whether multi-agent coordination adds genuine value net of transaction costs, and propose minimum evaluation standards as prerequisites for validating the CPH. 
\end{abstract}

\section{Introduction}\label{sec:introduction}

Financial markets reward decision quality under uncertainty. When LLM-based agents entered this domain, the initial approach treated the entire investment workflow as a single prompt-to-trade pipeline. These monolithic systems face a fundamental constraint: a single context window cannot simultaneously perform macroeconomic regime identification, earnings quality assessment, technical pattern recognition, risk budgeting, and execution optimization. The dominant paradigm now decomposes these tasks across cooperating agents, and systems such as FinCon~\cite{yu2024fincon}, TradingAgents~\cite{xiao2025tradingagents}, and HedgeAgents~\cite{li2025hedgeagents} report substantial gains, with claimed cumulative returns ranging from 23\% to over 400\%. However, the field has not established whether these gains reflect genuine design advances or artifacts of inconsistent evaluation methodology.

This survey addresses that gap. Rather than cataloguing systems or ranking them by reported performance, we ask a more tractable prior question: 
\textit{Given that cross-system comparisons are currently unreliable, what design choices are most worth investigating rigorously once evaluation standards improve?}
Our approach proceeds in four steps: a taxonomy decomposing the design space (Section~\ref{sec:taxonomy}), documentation of five systematic evaluation failures (Section~\ref{sec:evaluation}), the \textit{Coordination Primacy Hypothesis} (CPH) derived from structural patterns that survive those failures (Section~\ref{sec:coordination}), and the \textit{Coordination Breakeven Spread} (CBS) metric that operationalizes the hypothesis in deployment (Section~\ref{sec:tradeoffs}). Our contributions are analytical; we do not report new empirical results.

\section{Related Work}\label{sec:related}

Ding et al.~\cite{ding2024llm} survey LLM agents for financial trading but treat multi-agent coordination as one topic among many, without systematic comparison of coordination trade-offs. General multi-agent surveys~\cite{guo2024llm,talebirad2023multiagent} analyze communication protocols abstractly, without confronting domain-specific constraints such as coordination latency measurable in basis points of adverse price movement. Sun et al.~\cite{sun2024llm} survey LLM-based multi-agent reinforcement learning; no published financial system has successfully integrated LLM agents with formal optimization guarantees. FinCon's verbal reward function is a step toward structured decision optimization but lacks formal convergence guarantees.

Our work differs in two respects. We treat financial multi-agent systems (MAS) as decision architectures rather than software architectures, so every design description is accompanied by its decision-quality implication. We also foreground evaluation methodology as a first-class concern: the gap between reported and robust performance is large enough to affect whether published claims should be acted upon.

\section{A Taxonomy of Design Patterns}\label{sec:taxonomy}

\subsection{System Selection}

Candidate systems were drawn from the LLM-based financial trading literature published between 2023 and 2026. Inclusion required that a system employ distinct LLM roles for active trading or portfolio allocation and provide sufficient architectural detail for four-dimensional classification. Purely single-agent systems, rule-based pipelines, and systems lacking architectural documentation were excluded.
These criteria yield 12 systems, selected to maximize coverage across the four taxonomy dimensions: FinCon~\cite{yu2024fincon}, TradingAgents~\cite{xiao2025tradingagents}, HedgeAgents~\cite{li2025hedgeagents}, ContestTrade~\cite{zhao2025contesttrade}, FinVision~\cite{fatemi2024finvision}, TradingGPT~\cite{li2023tradinggpt}, QuantAgents~\cite{li2025quantagents}, AlphaAgents~\cite{blackrock2025alphaagents}, ElliottAgents~\cite{wawer2025elliottagents}, FinRobot~\cite{yang2024finrobot}, Agentic RAG~\cite{cook2025agentic}, and Chinese Public REITs system~\cite{li2026reits}. We include FinMem~\cite{yu2024finmem} (memory-focused) and FinAgent~\cite{zhang2024finagent} (tool-augmented) as single-agent baselines for comparison in Table~\ref{tab:performance}.

\subsection{Four Dimensions}

\begin{figure}[t]
\centering
\includegraphics[width=\textwidth]{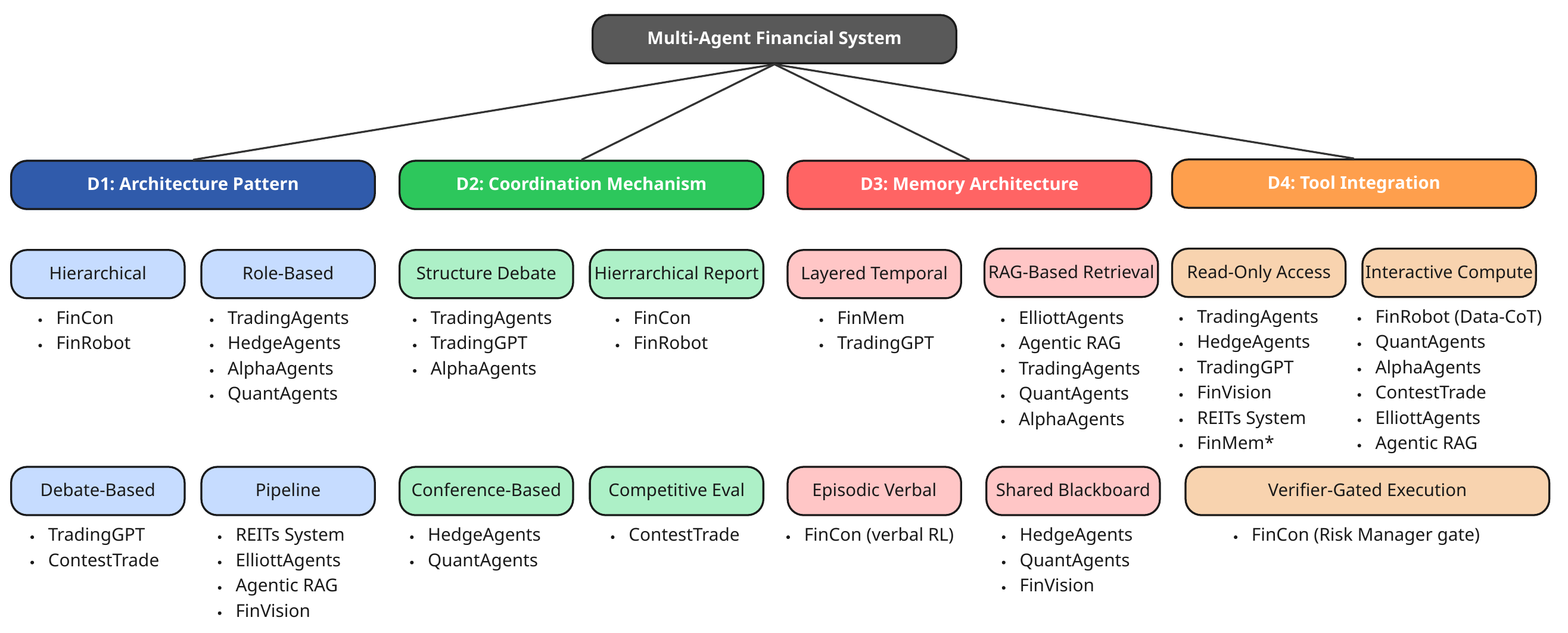}
\caption{Taxonomy of LLM-based multi-agent financial systems across four design dimensions. FinVision, ElliottAgents, REITs System, and Agentic RAG are not listed in Coordination Mechanism because they simply utilize a sequential pass-through (minimal
coordination).}

\label{fig:taxonomy}
\end{figure}

We classify systems along four dimensions (Fig.~\ref{fig:taxonomy}), intended to decompose hybrid designs so that future controlled experiments can vary one dimension while holding others constant.

\subsubsection{D1: Architecture Pattern.}
\textit{Hierarchical} architectures use a manager agent to arbitrate specialist inputs; the manager's ability to weight conflicting inputs is the binding constraint. \textit{Role-based} designs map agents to professional departments, supporting auditability. \textit{Debate-based} architectures improve calibration under signal ambiguity but incur overhead costly when execution speed matters. \textit{Pipeline} architectures offer low latency but no error correction.

\subsubsection{D2: Coordination Mechanism.}
\textit{Structured debate} improves accuracy over two to four rounds~\cite{du2023improving,choi2025debate} but risks Degeneration-of-Thought~\cite{liang2024degeneration}, where agents converge to a shared wrong answer
through social pressure. \textit{Hierarchical reporting} uses selective knowledge propagation to reduce noise, ensuring only decision-relevant feedback reaches specialists. \textit{Conference-based} coordination activates group discussions adaptively but requires precise triggers to avoid activating complex
protocols during routine trading. \textit{Competitive evaluation} rewards contrarian accuracy rather than consensus, avoiding consensus bias entirely. 
The absence of competitive risk-adjusted returns among systems using a sequential pass-through provides indirect motivating evidence for the CPH (Section~\ref{subsec:cph}),
though this observation is subject to the same evaluation confounds documented in Section~\ref{sec:evaluation} and should be treated as motivating rather than evidential.

\subsubsection{D3: Memory Architecture.}
\textit{Layered temporal} memory risks assuming fixed relevance decay during structural breaks. \textit{RAG-based retrieval} allows for high-granularity data access, but introduces experience-following behaviour~\cite{xiong2025memory}, amplifying anchoring bias. \textit{Episodic verbal} memory supports compliance and auditability but risks update lag. \textit{Shared blackboard} state enables real-time sharing but propagates errors system-wide.

\subsubsection{D4: Tool Integration.}
\textit{Read-only access} depends on LLM numerical reasoning, a documented weakness. \textit{Interactive computation} addresses this but introduces code correctness as a failure mode. \textit{Verifier-gated execution} validates outputs before action and is preferred for institutional deployment.

\subsection{Cross-System Observations}
\label{subsec:crosssystem}

Table~\ref{tab:performance} reports metrics for the eight systems that publish them; the remaining six (TradingGPT, AlphaAgents, ElliottAgents, FinRobot, REITs System, Agentic RAG) are framework papers without standardized trading metrics. The evaluation quality assessment scores each system against the five criteria introduced in Section~\ref{sec:evaluation}: no system satisfies more than two criteria. These eight systems differ along at least five methodological dimensions simultaneously: evaluation period, asset universe, market regime, cost model, and baseline choice. Normalizing across these dimensions would produce numbers that appear comparable but conceal the differences that matter most for practitioners. Instead, we assess how trustworthy each system’s reported numbers are (Evaluation Quality column) and section~\ref{sec:evaluation} addresses the evaluation inconsistencies in depth.

Although the results in Table~\ref{tab:performance} are not directly comparable, a qualitative structural pattern is worth noting.
Systems with explicit coordination mechanisms more often report extended evaluation horizons and live deployment, for example HedgeAgents posts a 405\% three-year return and QuantAgents sustains 1.76–2.02 live Sharpe (both subject to the evaluation quality limitations in Section~\ref{sec:evaluation}), whereas pipeline designs rarely disclose comparable long-horizon, risk-adjusted results. High Sharpe ratios observed over short, bullish windows (e.g., debate-based TradingAgents evaluations) highlight the importance of temporal robustness rather than establishing superiority. While not causal evidence, this pattern suggests coordination complexity may correlate with demonstrated robustness, motivating the CPH (Section~\ref{sec:coordination}).

\begin{table}[h]
\caption{
Reported performance metrics. Evaluation quality scores each system against five criteria:
(1)~contamination control, (2)~point-in-time universe, (3)~rolling-window reporting, (4)~net-of-cost returns, (5)~regime coverage.
$\checkmark$~=~satisfied, $\times$~=~not satisfied. 
$\dagger$~FinMem's reported 23\% return on MSFT reversed to $-$22\% under FINSABER controlled conditions.
}
\label{tab:performance}
{\centering
\renewcommand{\arraystretch}{1.3}
\setlength{\tabcolsep}{3pt}
\footnotesize
\resizebox{\textwidth}{!}{%
\begin{tabular}{|p{3.2cm}|c|c|c|c|c|p{3.5cm}|
  >{\centering\arraybackslash}p{0.5cm}
  >{\centering\arraybackslash}p{0.5cm}
  >{\centering\arraybackslash}p{0.5cm}
  >{\centering\arraybackslash}p{0.5cm}
  >{\centering\arraybackslash}p{0.5cm}|}
\hline
\rowcolor{gray!15}
\textbf{System \newline (Agents)}
  & \textbf{Sharpe Ratio}
  & \textbf{Cumulative Return}
  & \textbf{Annual Return}
  & \textbf{Max Drawdown}
  & \textbf{Eval Period}
  & \textbf{Evaluation Details}
  & \multicolumn{5}{c|}{\textbf{Evaluation Quality}} \\
\rowcolor{gray!15}
  & & & & & & &
  \rotatebox{90}{\scriptsize\textbf{Contam.}} &
  \rotatebox{90}{\scriptsize\textbf{Point-in-Time}} &
  \rotatebox{90}{\scriptsize\textbf{Rolling Win.}} &
  \rotatebox{90}{\scriptsize\textbf{Net Costs}} &
  \rotatebox{90}{\scriptsize\textbf{Regime Cov.}} \\
\hline
\textbf{FinAgent}~\cite{zhang2024finagent} \newline (1+tools)
  & 1.43--2.01 & --- & 31.9--92.3\% & 5.57--13.2\% & $\sim$6 mo
  & Six datasets; stocks and crypto. Best ARR 92.27\% on TSLA.
  & \yes & \no & \no & \no & \yes \\
  & & & & & & & \multicolumn{5}{c|}{\colorbox{yellow!25}{\textcolor{scoreamber}{$\blacksquare$}} \scoremid{2/5}} \\
\hline
\textbf{FinCon}~\cite{yu2024fincon} \newline (7+1)
  & 3.26 & 114\% & --- & 16.2\% & $\sim$18 mo
  & Jan 2022--Jun 2023; 6 US stocks
  & \yes & \no & \no & \no & \yes \\
  & & & & & & & \multicolumn{5}{c|}{\colorbox{yellow!25}{\textcolor{scoreamber}{$\blacksquare$}} \scoremid{2/5}} \\
\hline
\textbf{HedgeAgents}~\cite{li2025hedgeagents} \newline (4+1)
  & 2.41 & 405\% & 71.60\% & $\sim$14\% & 3 yr
  & 2021--2023; BTC, DJ30, Forex; multi-asset
  & \yes & \no & \no & \no & \yes \\
  & & & & & & & \multicolumn{5}{c|}{\colorbox{yellow!25}{\textcolor{scoreamber}{$\blacksquare$}} \scoremid{2/5}} \\
\hline
\textbf{QuantAgents}~\cite{li2025quantagents} \newline (Multi)
  & 3.11 & $\sim$300\% & 58.7\% & 16.86\% & 3 yr+live
  & Jan 2021--Dec 2023; NASDAQ-100.
  & \yes & \no & \no & \no & \yes \\
\cline{2-7}
  & 1.76-2.02 & 98-112\% & --- & --- & Q3 2024 - Q1 2025 & Live trading A-stock and HK-stock markets & \multicolumn{5}{c|}{\colorbox{yellow!25}{\textcolor{scoreamber}{$\blacksquare$}} \scoremid{2/5}} \\
\hline

\textbf{ContestTrade}~\cite{zhao2025contesttrade} \newline (Multi)
  & 3.12 & 52.8\% & --- & 12.41\% & ---
  & NASDAQ-100
  & \yes & \no & \no & \no & \no \\
  & & & & & & & \multicolumn{5}{c|}{\colorbox{red!15}{\textcolor{scorered}{$\blacksquare$}} \scorelow{1/5}} \\
\hline
\textbf{FinVision}~\cite{fatemi2024finvision} \newline (4)
  & 1.20--1.72 & --- & 14.8--42.1\% & 12.09--14.38\% & $\sim$7 mo
  & AAPL, MSFT, AMZN; predominantly bullish window; pipeline architecture.
  & \yes & \no & \no & \no & \no \\
  & & & & & & & \multicolumn{5}{c|}{\colorbox{red!15}{\textcolor{scorered}{$\blacksquare$}} \scorelow{1/5}} \\
\hline
\textbf{TradingAgents}~\cite{xiao2025tradingagents} \newline (7)
  & \textcolor{scoreamber}{5.60--8.21}
  & 23--27\% & 24.9--30.5\% & 0.91--2.1\% & 3 mo
  & Jan--Mar 2024; AAPL, GOOGL, AMZN. Sharpe ratio inflated by short bullish window.
  & \yes & \no & \no & \no & \no \\
  & & & & & & & \multicolumn{5}{c|}{\colorbox{red!15}{\textcolor{scorered}{$\blacksquare$}} \scorelow{1/5}} \\
\hline

\textbf{FinMem}~\cite{yu2024finmem}~$\dagger$ \newline (1)
  & 0.23--2.67
  & 23-61.7\%
  & ---
  & 10.8--22.9\%
  & $\sim$1 yr
  & Oct 2022-Apr 2023
  & \cellcolor{red!10}\textcolor{red!70!black}{\textbf{$\times$}}
  & \cellcolor{red!10}\textcolor{red!70!black}{\textbf{$\times$}}
  & \cellcolor{red!10}\textcolor{red!70!black}{\textbf{$\times$}}
  & \cellcolor{red!10}\textcolor{red!70!black}{\textbf{$\times$}}
  & \cellcolor{red!10}\textcolor{red!70!black}{\textbf{$\times$}} \\
\cline{2-7}
  & 1.4 $\to$ -1.24 & \textcolor{scorered}{\textbf{23$\to$ $-$22\%}} & &14.9 $\to$ -29 &6 mo $\to$ 8 mo  & MSFT only, Under FINSABER controlled re-evaluation. & \multicolumn{5}{c|}{\colorbox{red!15}{\textcolor{scorered}{$\blacksquare$}} \scorelow{0/5}} \\
\hline
\multicolumn{12}{l}{%
  \scriptsize
  \colorbox{green!15}{\textcolor{green!50!black}{$\blacksquare$}}~4--5/5 Relatively credible\quad
  \colorbox{yellow!25}{\textcolor{scoreamber}{$\blacksquare$}}~2--3/5 Partial credibility\quad
  \colorbox{red!15}{\textcolor{scorered}{$\blacksquare$}}~0--1/5 Low credibility
} \\
\end{tabular}
}
\par}
\normalsize
\end{table}

\begin{tcolorbox}[
  colback=warnbg, colframe=warnborder, boxrule=0.6pt, arc=3pt,
  left=6pt, right=6pt, top=4pt, bottom=4pt, before upper={\small}
]
\textbf{Note:} Metrics across rows are \textbf{not directly comparable} owing to
differences in evaluation period, asset universe, market regime, and cost assumptions.
\end{tcolorbox}

\section{Evaluation Failures in the Published Literature}\label{sec:evaluation}

The design diversity in Section~\ref{sec:taxonomy} means that any observed performance difference between two systems could reflect a genuine design advantage, a difference in evaluation conditions, or both. Five systematic failures prevent these explanations from being distinguished, making cross-system comparison unreliable as a basis for design conclusions.

\subsection{Look-Ahead Bias}\label{subsec:lookahead}
LLMs trained through 2024 may have encountered financial outcomes for periods used in backtesting, effectively retrieving rather than predicting. StockBench~\cite{chen2025stockbench} addresses this with DJIA data from March to July 2025; most LLM-based agents fail to outperform buy-and-hold under these conditions. A second manifestation is feature leakage through retrieval: imprecisely timestamped RAG databases can inject future information into historical queries. FinAgent's multi-step retrieval and Agentic RAG's cross-encoder re-ranking are both vulnerable in the absence of documented timestamp controls. 

\subsection{Survivorship Bias}\label{subsec:survivorship}
Most systems evaluate on stock universes selected at evaluation time, excluding delisted companies that are disproportionately poor performers. Elton et al.~\cite{elton1996survivorship} estimated 0.9\% annual survivorship bias in mutual fund returns; for individual stock selection the effect is larger. FINSABER~\cite{li2025finsaber} addresses this with historical index constituent lists. 

\subsection{Backtesting Overfitting}\label{subsec:overfitting}
LLM-based multi-agent systems have extensive hyperparameters (agent count, debate rounds, memory depth, temperature, prompt templates, evaluation windows), creating combinatorial space prone to overfitting. FinMem's reported 23.26\% cumulative return on MSFT became $-$22.04\% under a slightly different but equally defensible window with transaction costs included~\cite{li2025finsaber}. A sign reversal of this magnitude is consistent with an overfitted system.

\subsection{Transaction Cost Neglect}\label{subsec:txcost}
Round-trip costs of 10 to 20 basis points can compound to 25 to 50 percentage points of annual drag for daily-trading systems. Of systems surveyed, only FINSABER and StockBench explicitly model transaction costs. HedgeAgents' 405\%, FinCon's 114\%, and ContestTrade's 52.8\% are all gross of costs. This failure is particularly consequential for MAS: coordination-driven signal improvements may increase trading frequency without proportionally improving per-trade alpha, converting a nominal performance advantage into net underperformance. 

\subsection{Regime-Shift Blindness}\label{subsec:regime}
Most evaluations cover six to twelve months within a single market regime, providing no cross-regime evidence. Only HedgeAgents explicitly addresses regime adaptation; its three-year evaluation spanning 2021--2023 is the strongest available cross-regime evidence among surveyed systems. TradingAgents reports an extraordinary Sharpe of 5.60 to 8.21 based solely on a three-month bullish window (January–March 2024) during rallies in AAPL, GOOGL, and AMZN. 
A Sharpe ratio at this level, annualized from a single favourable regime, is statistically consistent with trend following in a favorable regime rather than genuine risk-adjusted alpha, and is consistent with regime shift blindness producing unreliable metrics.

\subsection{Consolidated Minimum Standards}\label{subsec:standards}
\begin{enumerate}
\item \textbf{Contamination control.} Evaluation period should post-date model training, or a post-training ablation should be provided.
\item \textbf{Point-in-time universe.} Asset universe should reflect historical index composition at each evaluation date.
\item \textbf{Rolling-window reporting.} Performance across multiple non-overlapping windows with variance estimates.
\item \textbf{Net-of-cost returns.} Explicit transaction cost model covering commissions, half-spread, and market impact.
\item \textbf{Regime coverage.} Evaluation spanning multiple regimes or explicit adversarial stress testing.
\end{enumerate}
No system in our survey satisfies all five (see evaluation quality scores in Table~\ref{tab:performance}).
FinCon, HedgeAgents, FinAgent, and QuantAgents each satisfy only 2/5, while TradingAgents, ContestTrade, and FinVision reach just 1/5. 
Notably, FinMem scores 0/5, with its reported 23\% return on MSFT reversing to -22\% under controlled re-evaluation. Building a benchmark satisfying all five simultaneously is among the most pressing infrastructure needs in this field.

\section{The Coordination Primacy Hypothesis}\label{sec:coordination}

\subsection{Motivation}
The evaluation failures above make precise quantitative comparison unreliable, but structural observations remain informative even when specific return figures do not: which systems survive the transition to live trading, which coordination patterns appear consistently across independent research groups, and which designs collapse under controlled re-evaluation. The CPH is derived from these patterns rather than from cross-system performance rankings.

\subsection{Hypothesis Statement}\label{subsec:cph}
\textit{Coordination Primacy Hypothesis holds that the inter-agent coordination protocol is the most consequential structural factor in trading decision quality among the four taxonomy dimensions, exerting greater influence than  model selection.}

This is a falsifiable claim: upgrading the LLM backbone within a fixed coordination protocol should yield smaller performance improvements than replacing the coordination protocol, holding all other design choices constant. The hypothesis does not assert that coordination is sufficient; only that it is the most consequential dimension to optimize.

\subsection{Supporting Evidence (Tiered)}

\subsubsection{Tier 1 -- Live-Market Benchmarking (Strongest).}
Available evidence suggests that framework architecture is a more consequential predictor of profitability than model selection: weaker models within sophisticated coordination structures tend to outperform frontier models in linear pipelines across the benchmarks examined. This is the most credible available evidence for the CPH, though regime diversity covered by AMA~\cite{qian2025ama} remains limited and the finding should be treated as strongly suggestive rather than definitive.

\subsubsection{Tier 2 -- Ablation Studies (Moderate).}
In FinCon and TradingAgents, removing the coordination reduced the Sharpe ratio by 15--30\%; while substituting a smaller model produced only 5--8\% variance. These ablations are author-reported and should be treated as suggestive rather than confirmatory.

\subsubsection{Tier 3 -- Theoretical Scaling Arguments (Tentative).}
Formal results~\cite{estornell2024multi} suggest that increasing agent count without an optimized coordination topology yields diminishing returns and increased inter-agent interference. This is consistent with the CPH but does not directly test it in financial settings.

\subsection{Why the CPH Cannot Yet Be Validated}\label{subsec:validation}
Definitive validation requires a controlled experiment varying D2 while holding D1, D3, D4, and LLM backbone constant, evaluated on contamination-free data with rolling windows and net-of-cost returns. This experiment has not been conducted because the five evaluation failures make its prerequisites unavailable. The failures are not merely a general critique; they are the specific obstacle blocking the field's most important untested hypothesis. Addressing them is a prerequisite for testing the CPH, not a parallel concern.

To transition toward empirical validation, we propose a Cross-Architecture Factorial Design. By isolating \emph{Coordination Logic} as the primary independent variable and \emph{LLM Parameter Scale} as a control variable, researchers can quantify the Marginal Alpha contributed by the protocol. We suggest a benchmark of 500 simulated trading days across three distinct market regimes (Bull, Bear, and Sideways) to ensure the coordination advantage is robust against regime-specific model biases.

\section{Coordination Trade-offs and the CBS Metric}\label{sec:tradeoffs}

If coordination protocol is the most consequential design dimension, understanding its costs and risks is essential for any practitioner acting on that hypothesis. This section synthesises four trade-off axes and introduces the CBS as the metric that operationalizes the CPH in deployment.

\subsection{Key Trade-off Axes}

\textbf{Cost and performance.} Inference costs scale linearly with agent count, while coordination costs scale quadratically in fully connected topologies. At current API pricing, a seven-agent system incurs roughly \$0.50--\$2.00 per daily decision, negligible for medium-frequency strategies but material at higher frequencies. Practical budgets of three to seven agents with two to three interaction rounds are consistent across the literature. Hybrid designs escalating selectively to frontier models for complex reasoning steps offer order-of-magnitude cost reductions.

\textbf{Debate and latency.} 
Each debate round introduces one to three seconds of latency; a two-round debate can incur five to twenty basis points of adverse price movement, potentially exceeding the signal improvement it provides. This latency cost is frequency-dependent, with coordination benefits most pronounced at medium-frequency horizons where holding periods are long enough to absorb the delay. Debate depth should be calibrated to both market conditions and asset type: direct execution during high-conviction signals minimises latency cost, while liquid equities can tolerate deeper coordination more readily than illiquid or volatile assets.

\textbf{Memory depth and regime drift.} Historical precedents from a prior regime introduce anchoring bias when structural conditions change. Existing designs such as FinMem's decay-based memory and FinCon's episodic updating partially address this but rely on fixed temporal structures rather than event-driven adaptation. Explicit regime-change detection is needed to trigger belief revision; absent such mechanisms, memory-equipped agents should incorporate circuit breakers suspending retrieval under elevated drawdown or volatility.

\textbf{Planner-executor depth.} Ablation results from FinCon and TradingAgents indicate that removing independent risk assessment degrades risk-adjusted performance. A minimal three-stage pipeline (signal generation, risk gate, execution) appears sufficient for most production settings, with additional stages yielding diminishing returns unless they contribute distinct information.

\subsection{The Coordination Breakeven Spread}\label{subsec:cbs}

The trade-off axes above share a common structure: coordination improves signal 
quality but incurs a cost, and no published metric determines whether the 
improvement justifies that cost for a given instrument. We formalize this via 
the \textbf{Coordination Breakeven Spread (CBS)}. Let $\Delta p(d)$ denote the 
expected improvement in entry/exit price from coordination depth $d$, and let 
$s$ denote the bid--ask spread (round-trip cost = $2s$). Defining

\[
\text{CBS}(d) = \frac{\Delta p(d)}{2}
\]

coordination is optimal only if $s < \text{CBS}(d)$; when the instrument's spread exceeds the CBS, coordination overhead is not recovered and the system should revert to single-agent operation.

Under reasonable assumptions, the implied CBS lies in the low single-digit basis-point range for typical Sharpe levels, indicating that coordination is economically viable primarily in highly liquid instruments with narrow spreads. 
In practice, $\Delta p(d)$ can be estimated as the difference in volume-weighted average execution price between a coordinated and single-agent baseline over matched trading windows, net of latency-induced slippage; where unavailable, it can be bounded using coordination-driven Sharpe improvements converted to basis points via average holding period. Precise calibration beyond these approximations remains an open empirical task.


\subsubsection{Relationship to the Evaluation Failures and the CPH.} CBS directly addresses transaction cost neglect (Section~\ref{subsec:txcost}) by converting coordination gains into a spread threshold. It is regime-dependent, as spreads widen during volatility spikes. In deployment, practitioners can test the CPH by running coordinated and single-agent systems in parallel and observing whether coordination consistently clears the CBS threshold.

\subsubsection{Asset-Class and Regime Dependence.} 
\begin{figure}[t]
\centering
\begin{tikzpicture}
\begin{axis}[
    width=1.0\textwidth, height=0.47\textwidth,
    xlabel={Instrument Bid-Ask Spread (bps)},
    ylabel={Coordination Signal Gain (bps)},
    xmin=0, xmax=80, ymin=0, ymax=80,
    xtick={0,20,40,60,80}, ytick={0,20,40,60,80},
    grid=both,
    grid style={line width=0.3pt, draw=gray!20},
    major grid style={line width=0.4pt, draw=gray!40},
    axis line style={gray!60}, tick style={gray!60}, font=\small,
]
\fill[green!8]  (axis cs:0,0) -- (axis cs:80,80) -- (axis cs:0,80) -- cycle;
\fill[red!6]    (axis cs:0,0) -- (axis cs:80,80) -- (axis cs:80,0) -- cycle;
\addplot[dashed, thick, color=blue!70, domain=0:80] {x};
\node[green!50!black, font=\small\bfseries] at (axis cs:15,72) {Coordination adds};
\node[green!50!black, font=\small\bfseries] at (axis cs:15,65) {genuine value};
\node[red!60!black, font=\small\bfseries]   at (axis cs:59,15) {Coordination destroys value};
\node[red!60!black, font=\footnotesize\itshape] at (axis cs:59,8) {(cost $>$ signal gain)};
\node[blue!70, rotate=23, font=\small\bfseries] at (axis cs:38,44) {CBS threshold};
\node[gray!70, font=\footnotesize, align=center] at (axis cs:10,30) {Large-cap\\equities};
\node[gray!70, font=\footnotesize, align=center] at (axis cs:5,50) {Forex\\majors};
\node[gray!70, font=\footnotesize, align=center] at (axis cs:30,15) {Small-cap\\equities};
\node[gray!70, font=\footnotesize, align=center] at (axis cs:68,52) {Crypto /\\EM instr.};
\end{axis}
\end{tikzpicture}
\caption{Conceptual illustration of the CBS threshold. Instruments in the upper-left region (low spread, high coordination gain) are candidates for multi-agent coordination; those in the lower-right are not. Asset-class regions are directional and illustrative only; empirical CBS values cannot be computed from currently published results owing to transaction cost neglect (Section~\ref{subsec:txcost}).}
\label{fig:cbs}
\end{figure}
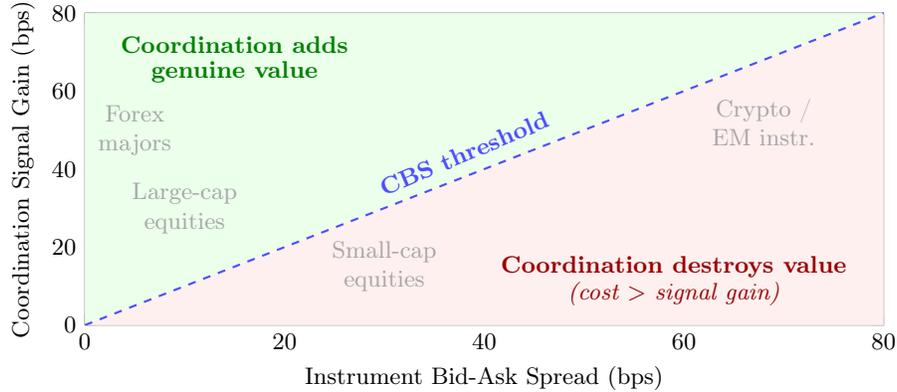

The CBS threshold varies substantially (Fig.~\ref{fig:cbs}). For large-capitalisation US equities (1--2 bps spreads), coordination-driven signal improvements may plausibly exceed costs at daily frequency. For small-capitalisation equities (10--50 bps), mid-capitalisation cryptocurrency (20--100 bps), or emerging market instruments, the threshold is considerably higher. A system applying coordination uniformly will over-coordinate during crisis periods when spreads are wide and under-coordinate during calm periods when coordination overhead is relatively more costly. We propose CBS as a standard reporting requirement alongside Sharpe ratio and maximum drawdown.

\section{Conclusion and Future Directions}\label{sec:conclusion}

This survey has argued that five systematic evaluation failures make cross-system comparisons of LLM-based multi-agent financial systems unreliable, and that addressing them is a prerequisite for any credible claim about what drives performance.
From the structural patterns that remain observable despite these failures, we formulated the CPH and introduced the CBS as its deployment metric. The logical chain is: the taxonomy provides vocabulary for controlled comparison; the evaluation critique explains why comparison is currently unreliable; the CPH identifies what is most worth testing; and the CBS defines what testing it requires in practice. We identify three high-priority directions for future work.

\textbf{Controlled validation of the CPH.} The definitive experiment holds architecture topology, memory design, tool integration, and LLM backbone constant while varying only the coordination mechanism across identical contamination-free data with rolling-window, net-of-cost evaluation. A community benchmark providing this infrastructure would either confirm the CPH and redirect research effort, or reveal that coordination and model quality require joint optimization.

\textbf{Small language model specialist architectures.} No published system implements a production-ready hybrid in which small models handle routine subtasks (sentiment classification, compliance checking) and escalate to frontier models only for complex reasoning. Related work~\cite{belcak2025small} suggests that fine-tuned small models can match frontier models on specialized tasks, with inference cost reductions of one to two orders of magnitude, though this finding derives from general agentic settings and its applicability to financial multi-agent systems remains an open question.

\textbf{Systemic risk from correlated AI trading.} If multiple institutions deploy similar LLM-based multi-agent architectures, correlated signals could amplify rather than dampen market volatility. Existing regulatory frameworks (IMF, CFTC, EU AI Act), calibrated for single-agent AI systems, do not account for emergent coordination effects across independent deployments.

The most consequential contributions will come not from adding agents or scaling models, but from designing coordination mechanisms that demonstrably improve risk-adjusted, net-of-cost, regime-robust decision quality, and from building the evaluation infrastructure needed to verify such claims.

\begin{credits}

\subsubsection{\discintname}
The authors declare no competing interests. No external funding was received in support of this work.
\end{credits}

\bibliographystyle{splncs04}
\bibliography{references}

\end{document}